\documentclass[preprint,aps,nofootinbib]{revtex4}
\usepackage{graphicx}
\usepackage{amsmath}
\usepackage{amsfonts}
\usepackage{amssymb}
\usepackage{color}%
\providecommand{\U}[1]{\protect\rule{.1in}{.1in}}

\newcommand{\f}{\begin{equation}}
\newcommand{\ff}{\end{equation}}
\newcommand{\fa}{\begin{eqnarray}}
\newcommand{\ffa}{\end{eqnarray}}

\begin{document}
\title{Interacting dark energy, holographic principle and coinindence problem}
\author{Bo Hu$^{1}$}
\email{bohu@ncu.edu.cn}
\author{Yi Ling$^{1,2}$}
\email{yling@ncu.edu.cn}
\affiliation{${}^{1}$ Center for Gravity and Relativistic Astrophysics, Department of
Physics, Nanchang University, 330047, China}
\affiliation{${}^{2}$ CCAST (World Laboratory), P.O. Box 8730, Beijing 100080, China}

\begin{abstract}
The interacting and holographic dark energy models involve two
important quantities. One is the characteristic size of the
holographic bound and the other is the coupling term of the
interaction between dark energy and dark matter. Rather than
fixing either of them, we present a detailed study of theoretical
relationships among these quantities and cosmological parameters
as well as observational constraints in a 
general formalism.
In particular, we argue that the ratio of dark matter to dark
energy density depends on the choice of these two quantities, thus
providing a mechanism to change the evolution history of the ratio
from that in standard cosmology such that the coincidence problem
may be solved. We investigate this problem in detail and construct explicit
models to demonstrate that
it may be alleviated provided that the interacting term and the characteristic
size of holographic bound are appropriately specified.
Furthermore, these models are well fitted with the current
observation at least in the low red-shift region.
\end{abstract}
\maketitle

\section{Introduction}

Recent astronomical observations indicate that our universe is
currently undergoing an epoch of accelerated expansion
\cite{Riess98cb,Perlmutter98np,Bennett03bz,Spergel03cb,Tegmark03ud,Tegmark03uf}%
. Following the standard Friedmann-Robertson-Walker (FRW)
cosmology such an expansion implies the existence of a dark energy
(DE) component to the mass-energy density of the
Universe\footnote{For recent discussion on the possibility of
constructing accelerating universe without dark energy, see for
instance \cite{Kolb05me,Kolb05da,Ishibashi05sj,Kolb05ze}.}. At
present it is fair to say that disclosing the nature of DE is one
of the central problems in the research of both cosmology and
theoretical physics (for recent reviews, see
\cite{Weinberg88cp,Carroll00fy,Peebles02gy}). In this direction we
are faced with many fundamental and difficult issues, among of
which the following two open questions are of particular
importance. The first one is on the nature and dynamical
properties of the dark energy. Currently it is not clear yet
whether DE can be described by cosmological constant which is
independent of time, or by dynamical fields such as quintessence,
K-essence, tachyon fields or phantom fields. The second is the
coincidence problem, dubbed as \textquotedblleft\ why are the
densities of matter and dark energy of precisely the same order
today?" \cite{coincidence}.

To shed light on these two open questions, some interesting DE
models were proposed recently. Those models can be divided into
two categories, i.e. the holographic dark energy (HDE) models and
the interacting DE models. The former stems from the holographic
hypothesis \cite{Cohen98zx,Hsu04ri,Li04rb} and can provide an intriguing way to 
interpret the dynamics of DE, while it is suggested that the latter can help to understand the coincidence problem by considering the possible
interaction between dark energy and cold dark matter
\cite{Amendola99er,Chimento03ie,Mangano02gg,Farrar03uw,Pavon05yx}.

Let us first start with a close look on the holographic dark energy model
motivated from the holographic hypothesis, which has gradually been believed
to be a fundamental principle in the quantum theory of gravity. According to
this principle, the number of degrees of freedom for a system within a finite
region should be finite and is bounded roughly by the area of its boundary.
While in a cosmological setting the challenge is to put a reasonable and
well-defined upper bound on the entropy of the universe. Motivated by
Bekenstein entropy bound, it seems plausible to require that for an effective
quantum field theory in a box of size $L$ with UV cutoff $\Lambda$, the total
entropy should satisfy the relation
\begin{equation}
L^{3}\Lambda^{3}\leq S_{BH}=\pi L^{2}M_{p}^{2},
\end{equation}
where $S_{BH}$ is the entropy of a black hole with the same size
$L$. but further consideration indicates that to saturate this
inequality some states with Schwarzschild radius much larger than
the box size have to be counted in. As a result a stronger entropy
bound has been proposed in \cite{Cohen98zx}, requiring that the
total energy of system with size $L$ should not exceed the mass of
a black hole with the same radius, namely
\begin{equation}
L^{3}\Lambda^{4}=L^{3}\rho_{\Lambda}\leq LM_{p}^{2}.
\end{equation}
While saturating this inequality by choosing the largest $L$ it gives rise to
a holographic energy density
\begin{equation}
\rho_{\Lambda}=3c^{2}M_{p}^{2}L^{-2},
\end{equation}
where $c$ is a dimensionless constant. Then the key issue is 
what possible physical scale one can choose as the cutoff $L$
 constrained by the fact of the current acceleration of the universe.
Originally the natural choice is to identify the Hubble horizon as
$L$, however, as pointed out in \cite{Hsu04ri}, this will lead to
a wrong equation of state for dark matter which conflicts with the
ordinary one in standard cosmology. As a result, in \cite{Li04rb}
Li proposed to take the future event horizon as the largest size
$L$, which gives rise to desired results and then stimulate 
a lot of interests and discussions 
in this subject\cite{holoref}. However, there are some
unsatisfactory points in this conjecture. First, it still remains
puzzling how the current evolution of dark energy density can be
determined by the \textit{future} event horizon. Second, the
coincidence problem can hardly 
be solved in this context.

As pointed out in \cite{Pavon05yx}, the reason that one is forced
to take the future event horizon is based on the assumption that
the energy densities of dark energy and dark matter evolve
independently. However, if there exists interaction between DE and
dark matter (DM), then the cutoff $L$ is not necessarily
identified as the future event horizon. As a matter of fact, the
interaction between DE and DM is proposed to solve the coincidence
problem and has been discussed in many recent works
\cite{interef}. This can be accomplished by introducing the
coupling terms in the equations of state for matter and dark
energy densities, which can bring the ratio of these two
ingredients into a constant at late times. From the theoretical
point of view this sort of coupling is completely possible due to
the unknown nature of DM and DE. In addition, this proposal is
compatible with the current observations such as the SNIa and WMAP
data\cite{Chimento03ie}, and even favored in some circumstances as
suggested in \cite{Szydlowski05ph}. But until now, only certain special interacting terms have been considered in 
existing literatures. 

Now based on the above discussion, it is natural to ask if we
could combine these two theoretical proposals 
together so as to improve our understanding 
of dark energy. This is the main purpose of our paper. Although
there are many existing works in both two directions, most
discussions in those works only considered 
specific characteristic sizes or interacting terms. For instance,
the characteristic size is usually assumed to be the future event
horizon after the work \cite{Li04rb} and the interaction term is
assumed to take a form as $3b^2H\rho$ where $b$ is a coupling constant.
However, there is no strong theoretical motivations for these
choices. 
In this paper, rather than fixing either the interacting term or
the holographic characteristic size, we intend to investigate the
nature of interacting and holographic DE  and the coincidence
problem in a more general formalism.

Our paper is organized as follows. We first present brief reviews
on interacting dark energy and holographic dark energy in Section
2 and Section 3, respectively. In particular, given the conditions
that our universe is currently accelerating and the ratio of dark
matter density to dark energy density deceases, we derive general
constraints on the relations among the interacting term,
holographic size and the equation of state parameter of dark
energy $\omega$. Then we turn to the coincidence problem in
Section 4, under the simplest requirement that the ratio of dark
matter to dark energy density to be constant. In Section 5 we
consider the case that the ratio can vary with time slowly and
demonstrate how the coincidence problem can be alleviated through
some specific examples by providing appropriate interacting terms
and holographic sizes.

\section{Interacting dark energy}

We start with the standard Friedmann equations in which DE and DM\
are assumed to be independent and there is no interaction between
them. Provided our current universe is dominated by dark energy
with the state equations $p=\omega\rho_{\Lambda}$ and cold dark
matter with $p=0$, these equations read as
\begin{equation}
\rho_{\Lambda}+\rho_{M}=3M_{p}^{2}H^{2},\label{f1}%
\end{equation}%
\begin{equation}
\dot{H}=-{\frac{1}{2}}M_{p}^{-2}[\rho_{\Lambda}(1+\omega)+\rho_{M}],\label{f2}%
\end{equation}%
\begin{equation}
\dot{\rho}_{\Lambda}+\dot{\rho}_{M}=-3H[\rho_{\Lambda}(1+\omega)+\rho
_{M}],\label{f3}%
\end{equation}
where $H=\dot{a}/a$ is the Hubble factor. It is well known that equations
(\ref{f1})-(\ref{f3}) are not independent and any one of them can be derived
from the other two. Introduce $\Omega_{\Lambda}=\rho_{\Lambda}/(3M_{p}%
^{2}H^{2})$ and $\Omega_{M}=\rho_{M}/(3M_{p}^{2}H^{2})$, the first Friedmann
can also be written as $\Omega_{\Lambda}+\Omega_{M}=1$.

Now we proceed to interacting dark energy models in which dark
matter and dark energy are postulated to be coupled such that dark
energy can decay into cold dark matter. As a result, the last
equation can be written as the combination of following two
evolving equations,
\begin{equation}
\dot{\rho}_{\Lambda}=-3H\rho_{\Lambda}(1+\omega)-Q,\label{i1}%
\end{equation}%
\begin{equation}
\dot{\rho}_{M}=-3H\rho_{M}+Q,\label{i2}%
\end{equation}
where $Q$ denotes the interacting term. 
To be a realistic model, the interacting DE model should satisfy
the observational constraints. 
First, we
consider the constraint on $Q$ by the observation that our
universe is currently accelerating. Since the ratio of dark matter
to dark energy plays a special role and its dynamics is a major
subject of this paper, for convenience we denote the ratio
$\rho_{M}/\rho_{\Lambda}$ by $r$ which is related to
$\Omega_{\Lambda}$ by $1+r=1/\Omega_{\Lambda}$. 
Then from (\ref{f1}) and (\ref{f2}) we have
\begin{equation}
\dot{H}=-{\frac{3}{2}}(1+{\frac{\omega}{{1+r}}})H^{2}.\label{f4}%
\end{equation}
Noticed that equation (\ref{f4}) always holds no matter whether
the interaction is taken into account or not. The solution to this
equation can be formally written as
\begin{equation}
H=H_{0}e^{-{\frac{3}{2}}\int_{0}^{x}\left(  1+{\frac{\omega}{1+r}}\right)
dx},\label{f4a}%
\end{equation}
where $x\equiv \ln a$. As a result, the requirement $\ddot{a}>0$
leads to
\begin{equation}
1+r+3\omega<0.\label{h2}%
\end{equation}
On the other hand, from (\ref{i1}) and (\ref{i2}) we find that the
interacting term has the following general form,
\begin{equation}
\tilde{Q}\equiv{\frac{Q}{H\rho_{\Lambda}}}={\frac{\dot{r}}{{(1+r)H}}}%
-{\frac{r}{1+r}}3\omega.\label{h3}%
\end{equation}
where $H$ can be absorbed by redefining $\dot{r}/H=dr/dx\equiv
r^{\prime},$ such that
\begin{equation}
\tilde{Q}={\frac{1}{1+r}}(r^{\prime}-3\omega r).\label{i30}%
\end{equation}
Then from (\ref{h2}) one finds that
\begin{equation}
\tilde{Q}>r+{\frac{r^{\prime}}{1+r}}.\label{i3}%
\end{equation}

Furthermore, it is expected that the ratio of dark matter density
to dark energy density decreases with the evolution of the
universe, namely, $r^{\prime}<0$. This requirement together with
the previous one (\ref{h2}) implies that the interacting term
should satisfy the following constraint
\begin{equation}
r+{\frac{r^{\prime}}{1+r}}<\tilde{Q}<{\frac{-3\omega r}{1+r}}.\label{i4}%
\end{equation}

\section{Holographic dark energy}

Now we turn to the holographic dark energy (HDE) models. As
introduced in Section 1, in this context the dark energy density
is assumed to be saturated in the region with size $L$,
\begin{equation}
\rho_{\Lambda}=3c^{2}M_{p}^{2}L^{-2}.
\end{equation}
Comparing this bound with the first Friedmann equation we easily
obtain a relation between the characteristic size $L$ and the
Hubble factor $H$ as
\begin{equation}
LH=\sqrt{1+r}c.\label{c1}%
\end{equation}

In general the characteristic size $L$ need be chosen in such a
way that the dark energy can be responsible for the acceleration
of the universe. Furthermore, such choice should not conflict with
the state equations of dark energy and dark matter. As a result,
when there is no interaction between DE\ and DM, i.e.
$\tilde{Q}=0$, $L$ is conventionally taken as the future event
horizon \cite{Li04rb,holoref} so as to fit the observational data.
However, as explained in Section 5, 
the coincidence problem
is hardly solved by pure HDE alone. Therefore, in the following
discussion
on HDE models, we will consider the case with non-vanishing $\tilde{Q}%
$. Moreover, instead of fixing $L$ or the interacting term
$\tilde{Q}$ at the beginning, 
we will take a more
phenomenological view and consider them to be free dynamical
quantities but constrained by observations.

Now, as in the previous section, from the requirements of
accelerating universe and decreasing ratio of DM density to DE
density one can derive the constraints on those quantities. Taking
the derivative with respect to time on both sides of equation
(\ref{c1}) and using (\ref{i30}) lead to a relation between $L$
and $\tilde{Q}$ as
\begin{equation}
\tilde{Q}=r^{\prime}-2r({\frac{L^{\prime}}{L}}-{\frac{3}{2}}).\label{qo}%
\end{equation}
Now due to the constraint (\ref{i4}) we find the size $L$ should satisfy
\begin{equation}
{\frac{3}{2}}({\frac{r^{\prime}}{3r}}+{\frac{\omega}{1+r}}+1)<{\frac
{L^{\prime}}{L}}<{\frac{r^{\prime}}{2(1+r)}}+1.\label{h4}%
\end{equation}
Thus we find in a general formalism of interacting holographic
dark energy, the interacting term and the characteristic size of
holographic bound are constrained by the inequalities (\ref{i4})
and (\ref{h4}), respectively.

Alternatively from (\ref{i30}) and (\ref{qo}) we may have a relation between
$L$ and $\omega$ as
\begin{equation}
2{\frac{L^{\prime}}{L}}=3(1+{\frac{\omega}{1+r}})+{\frac{r^{\prime}}{1+r}}.
\label{lo}%
\end{equation}
Thus $r^{\prime}<0$ leads to
\begin{equation}
{\frac{L^{\prime}}{L}}<{\frac{3}{2}}(1+{\frac{\omega}{1+r}}).
\end{equation}

Furthermore replacing the parameter $\omega$ appearing in (\ref{i4}) by $L$ we
may find the following inequality
\begin{equation}
-2r({\frac{L^{\prime}}{L}}-{\frac{3}{2}})>\tilde{Q}>2({\frac{L^{^{\prime}}}%
{L}}+{\frac{r}{2}}-1).
\end{equation}

In summary, we find for interacting holographic dark energy
models, $L$, $\tilde{Q}$, $\omega$ and $r$ are not independent
quantities but related by equations (\ref{qo}) and (\ref{lo}), or
(\ref{i30}) and (\ref{lo}) since among these three equations only
two of them are independent. Thus given any two of them 
then the dyanamics of the other two can be determined. For
instance, if we specify the interaction term $\tilde{Q}$ and the
characteristic size $L$, then the dynamics of $\omega$ and $r$ may
be determined and vice versa. However if only one of them is
specified, then the dynamics of the other three can not be
uniquely fixed. For example, in an interacting HDE model with the
future event horizon as the characteristic scale $L$, i.e.
\begin{equation}
L=a(t)\int_{t}^{\infty}{\frac{1}{a(t^{\prime})}}dt^{\prime},
\end{equation}
which gives rise to a relation $\dot{L}=LH-1$. Then from (\ref{lo}) it is easy
to derive the following relation
\begin{equation}
{\frac{1}{2}}(1+r+3\omega)+{\frac{\sqrt{1+r}}{c}}+{\frac{r^{\prime}}{2}%
}=0.\label{h1}%
\end{equation}
In this case $r^{\prime}<0$ requires that
\begin{equation}
1+r+3\omega>-{\frac{2\sqrt{1+r}}{c}}.\label{h5}%
\end{equation}
If we further specify the interaction term, e.g.
$\tilde{Q}=3b^{2}(1+r)$ 
as in references\cite{interef}, then the dynamics of $\omega$ and
$r$ can be determined uniquely, as discussed 
in \cite{Wang05jx}. However, since our goal is to
investigate the coincidence problem, we intend to put constraints
on the evolution of $r$ and then explore what expressions the
other quantities including $L$,$\tilde{Q}$ and $\omega$ may take.
This is what we are going to do in next two sections.

\section{Coincidence}

Before proceed, 
we first demonstrate how the coincidence problem arises in the standard cosmology. 
Setting $Q=0$ in (\ref{i1}) and
(\ref{i2}) will lead to $r^{\prime}=3\omega r$ (see also
(\ref{i30})). Now from (\ref{h2}) one finds that%
\[
\frac{d\ln r}{dx}=3\omega<-r-1<-1,
\]
which means during acceleration $r$ decreases faster than $a^{-1}$. 
Furthermore, from Friedmann equations,
one finds that
\begin{equation}
r=r_{0}\left(  a/a_{0}\right)  ^{-3},\label{k1}%
\end{equation}
and%
\begin{equation}
a/a_{0}=C\left(  e^{\lambda t/t_{0}}-e^{-\lambda t/t_{0}}\right)
^{2/3},\label{k2}%
\end{equation}
where $C$ and $\lambda$ are $O(1)$ constants which can be related
to the current values of $\Omega_{\Lambda}$. Then it is easy to
see 
when $t\ll t_{0}$, $r\propto t^{-2}$ and $r$ decreases
quadratically as expected for a matter dominated universe and when
$t\gg t_{0}$, $r$ deceases exponentially as expected in a dark
energy dominated universe. Then it is only when $t$ is around
$t_{0}$ that $r\thicksim O(1)$.

It is expected that adding interaction may change the dynamics of
$r$ greatly. In this section we consider the simplest possibility
with $\dot{r}=0$, which implies
$\rho_{\Lambda}\propto\rho_{M}\propto H^{2}$. It is worthwhile to
stress that this situation only occurs at late times. 
Suppose the ratio $r$ is a constant, i.e. $r=r_{0}$. We
immediately obtain the following equations
\begin{equation}
\tilde{Q}={\frac{-3\omega r_{0}}{1+r_{0}}},\label{i4c}%
\end{equation}%
\begin{equation}
\dot{L}={\frac{3c}{2}}\sqrt{1+r_{0}}(1+{\frac{\omega}{{1+r_{0}}}}).\label{f5c}%
\end{equation}
In addition, the Hubble factor is inversely proportional to
characteristic size $L$ as $H=\sqrt{1+r_{0}}cL^{-1}$. Therefore
specifying any one of quantities $\tilde{Q},L,\omega$, the
dynamics
of the other two 
can be uniquely determined from above equations. We classify some
possibilities in the following subsections.

\subsection{$\dot{r}=0$ with only interaction term specified}

One possible choice for the interaction term is 
setting $\tilde {Q}=3b^{2}(1+r_{0})$ as in previous references,
where $b$ is a constant. Then from (\ref{i4c}) we find that
$\omega$ is fixed as
\begin{equation}
\omega=-b^{2}\frac{(1+r_{0})^{2}}{r_{0}}.\label{j5}%
\end{equation}
Consequently the solutions to $H$ and $L$ can be obtained from (\ref{f5c}) as
\begin{equation}
H=H_{0}a^{-{\frac{3}{2}}(1-b^{2}-{\frac{b^{2}}{r_{0}}})}.
\end{equation}
In addition, from (\ref{i4}) one finds that
\begin{equation}
3b^{2}>\frac{r_{0}}{(1+r_{0})}=\Omega_{M}.\label{j6}%
\end{equation}
Therefore, it is obvious that in non-interacting models (i.e.
$b=0$) $\dot {r}=0$ and $\ddot{a}>0$ cannot be achieved
simultaneously. This can be considered as an important hint for
the need of interacting dark energy.

\subsection{$\dot{r}=0$ with only holographic characteristic size specified}

As shown in the previous section, specifying the holographic
characteristic size will determine $\tilde{Q}$ and $\omega$ since
$r$ has already been fixed to be $r_{0}$. 
Here we consider the HDE\ model with $L$ being the future event
horizon. From (\ref{i4c}) and (\ref{f5c}) we find correspondingly
that the parameter $\omega$ and $\tilde{Q}$ have to be constants
as well.
\begin{equation}
\tilde{Q}=\tilde{Q_{0}}=r_{0}(1+{\frac{2}{\sqrt{1+r_{0}}c}}).
\end{equation}
\begin{equation}
\omega=\omega_{0}=-{\frac{1}{3}}(1+r_{0}+{\frac{2\sqrt{1+r_{0}}}{c}}).\label{jj5}
\end{equation}

Using the current data $\rho_{M0}\simeq0.25$,
$\rho_{\Lambda0}\simeq0.72$ and setting $c=1$, we find
\begin{equation}
r_{0}\simeq0.35,\ \ \ \ \ \ \ \ \ \ \ \omega_{0}\simeq-1.22, \label{j7}%
\end{equation}
which is a phantom-preferred model. From (\ref{f4}) we find that
\begin{equation}
a\sim t^{\frac{2(1+r_{0})}{3(1+r_{0}+\omega_{0})}},
\end{equation}
while
\begin{equation}
\rho_{\Lambda}\sim\rho_{M}\sim t^{-2}.
\end{equation}

\subsection{$\dot{r}=0$ with both interaction term and holographic
characteristic size specified}

If both $L$ and $\tilde{Q}$ are specified as previous subsections,
then from (\ref{j5}) and (\ref{jj5}) it is easy to see $\dot{r}=0$
can be reached only when the constant $b^{2}$ takes the value as
\begin{equation}
b^{2}={\frac{-\omega_{0}r_{0}}{(1+r_{0})^{2}}}\simeq0.24.
\end{equation}

\section{soft coincidence}

The above discussion shows that the interaction between DE and DM\
can lead to a constant $r$. 
Although it is not clear how to obtain a $r$ of $O(1)$ size at
early times, this simple strategy can be used to account for
particular situations (e.g. late time evolution of the universe).
Nevertheless, there is no strong motivation for setting $r$ to be
a constant.
It is worthwhile to explore some more realistic models in which 
$r$ varies slowly with time.  We discuss this possibility in detail here. 

Advocated by the above discussion, 
one expects that certain amount of interaction can 
alleviate the coincidence problem. One possibility which has been
proposed in \cite{Chimento03ie} is to allow the ratio of two
energy densities to vary slowly but require that there are two
positive solutions $r_{\pm}$ to $\dot{r}=0$. Then the coincidence
can be alleviated if $r_{-}$ is close to $O(1)$
as 
the ratio $r$ evolves from the unstable but finite maximum $r_+$
to a stable minimum $r_-$ at late time, instead of from $\infty$
to 0. To demonstrate this possibility we consider again the model
with an interaction
term $\tilde{Q}=3b^{2}(1+r)$ and a constant $\omega$. 
From equation (\ref{i30}), we have
\begin{equation}
r^{\prime}=3b^{2}(1+r)^{2}+3\omega r\label{r1}.
\end{equation}
Obviously setting $r^{\prime}=0$ the equation has two positive
solutions with a relation $r_{+}r_{-}=1$. It is also possible
to show that the ratio will run from an unstable but finite
maximum $r_{+}$ to a stable minimum $r_{-}$ at late
time\cite{Chimento03ie}.

However, this does not occur in the context of pure holographic
dark energy if one chooses the future event horizon as the
characteristic size $L$. In the absence of interaction, the
dynamics of $r$ is described by
\begin{equation}
r^{\prime}=-r(1+{\frac{2}{c\sqrt{1+r}}}).
\end{equation}
Defining $\sqrt{1+r}=y$ $(y\leq1)$ leads to
\begin{equation}
2cy^{2}y^{\prime}=(1-y^{2})(cy+2).\label{hy}
\end{equation}
There is only one positive solution to $r^{\prime}=0$ with $y=1$
or $r=0$. Thus, as in standard cosmology $r$ runs from infinity to
zero and consequently the coincidence problem still exits in this
setting.

Next we intend to propose an alternative way to alleviate the
coincidence problem. That is, it might not be necessary to have
both an unstable finite maximum and a stable minimum close to
$O(1)$. The later is more important in the coincidence problem and
presumably is determined by the physics effective at the current
evolution of the universe. The former is more related to the early
evolution of the universe and whether or not an $O(1)$ initial
condition can be obtained is determined by physics beyond the
scope of this work. Therefore, we would rather leave the question
concerning the existence of a positive maximum open and
concentrate on the models with a positive stable minimum at late
time. In particular, if we find this stable value is not quite far
from the current observation then the coincidence problem may be
alleviated as the universe has a long time to stay at this stage
with a similar ratio. Now as an example consider adding the
interaction term $\tilde{Q}=3b^{2}(1+r)$ into the holographic dark
energy model presented above. 
Now we find 
 equation (\ref{hy}) is changed to
\begin{equation}
cy^{2}y^{\prime}={\frac{c}{2}}(3b^{2}-1)y^{3}-y^{2}+{\frac{c}{2}%
}y+1.\label{r1a}
\end{equation}
Provided $c>0$ and $3b^2<1$, it still has only one positive
solution to $y^{\prime}=0$ but {\it not}
at $r=0$. 
The exact position of the minimum depends on the values of $c$ and
$b$. For explicitness we illustrate the evolution of $r$ in FIG.
\ref{fig3} with $c=1$ and $b^{2}=0.12$, which is described by the
dot-dashed curve.

Moreover, in the discussion of (\ref{r1}) from which two positive
solutions are obtained for $r^{\prime}=0$, we have made two
assumptions, i.e. the coefficients $b^{2}$ and $\omega$ are
constants. 
A time dependent $b^{2}$ or$\ \omega$ might change this situation.
In addition, as mentioned in section 3, specifying any two of $L$
(or $\rho_{\Lambda}$), $\tilde{Q}$, $\omega$ and $r$ will
determine the other two uniquely. 
In principle,
$\omega$ 
might be different from those assumed in the discussion of
(\ref{r1}) and consequently will lead to different results. Below
we will discuss this in more detail through the following models.

\subsection{Model 1: Given $\tilde{Q}$ with time dependent $b^{2}$ }

In this model we assume that%
\[
b^{2}=b_{c}^{2}e^{-r/R},
\]
i.e.
\begin{equation}
\tilde{Q}=3b_{c}^{2}(1+r)e^{-r/R} \label{r2},
\end{equation}
where $b_{c}^{2}=$constant. The interaction given by (\ref{r2})
decreases exponentially as $r$ increases and consequently at early
times when $r\gg R$, $\tilde{Q}$ is very small and thus can be
ignored. Therefore, in this model the early age of the universe
can be described by the standard Friedmann equations without
interaction. The interaction becomes important only at late time
and will lead to a stable minimum which can mitigate the
coincidence
problem. As (\ref{i30}), one finds that in this case
\begin{equation}
r^{\prime}=3b_{c}^{2}e^{-r/R}(1+r)^{2}+3\omega r,\label{r3}
\end{equation}
and subsequently from (\ref{qo}) one have
\begin{equation}
{L'\over L}={3\over 2}[1+\omega+b_c^2(1+r)e^{-r/R}]\label{rr3}.
\end{equation}
Now to obtain more explicit results we have two options. One is to
set $\omega$ to be a constant, for example $\omega=-1$. This is
completely possible in the presence of holographic dark energy and
the corresponding size of holographic bound is determined by
\begin{equation}
{L'\over L}={3\over 2}b_c^2(1+r)e^{-r/R}.
\end{equation}
With this choice the equation (\ref{r3})has only one solution to
$r^{\prime}=0$ for small $R$. For example, when $b_{c}^{2}=0.3$,$\
R=0.25$ ,
the solution to $r^{\prime}=0$ is
$r_{f}=0.196.$
The second option, instead of specifying $\omega$, is setting an
appropriate characteristic size $L$ which can also lead to the
same results obtained above. 
To show this, we consider the HDE
model with $L$ being
the future horizon. From (\ref{rr3}) 
one has
\begin{equation}
3\omega=-1-3b_{c}^{2}(1+r)e^{-r/R}-\frac{2}{\sqrt{1+r}c}. \label{r4}
\end{equation}
which lead to
\begin{equation}
\frac{d\ln r}{dx}=\frac{r^{\prime}}{r}=-1+3b_{c}^{2}e^{-r/R}\frac{1+r}
{r}-\frac{2}{\sqrt{1+r}c}. \label{r5}
\end{equation}
For $b_{c}^{2}=0.3$, $R=0.5$ and $c=1$, the solution to
$r^{\prime}=0$ is $r_{f}=0.246$. In addition, one can check that
for $r>r_{f}$ one always has $\dot{r}<0$. In addition, in this
case $\omega$ also can across $-1$ as in many DE\ models. For
instance, for $b_{c}^{2}=0.1$, $c=1$ and $R=1$, $r_{f}=0.102$, one
finds that $\omega$ across $-1$ at $r=0.35$.

\subsection{Model 2: Given $\tilde{Q}$ with time independent $b^2$}

\begin{figure}
[ptb]
\begin{center}
\includegraphics[
height=3.7524in, width=6.0502in
]
{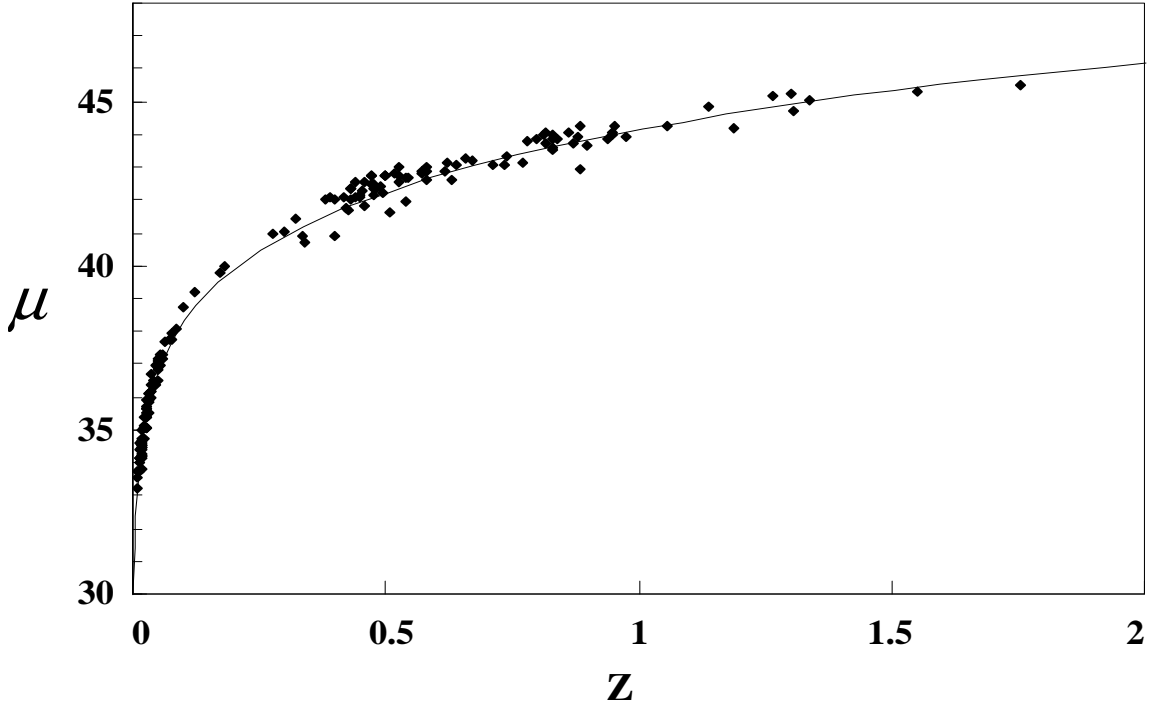}
\caption{Distance moduli vs. redshift plot for model 2
($b_{c}^{2}=0.12$, $r_{f}=0.2$). The data points are from the gold
sample of type Ia supernovae
of \cite{Riess98cb}.}
\label{fig1}
\end{center}
\end{figure}

In this subsection we will consider the situation where the
dynamics of $r$ at late time can be approximated by a power law
dependence on $a$, i.e.
\begin{equation}
r=r_{f}+(r_{0}-r_{f})a^{-k}, \label{r6}%
\end{equation}under the
assumption that $\tilde{Q}=3b_{c}^{2}(1+r)$ where $k=$
$3b_{c}^{2}/r_{f}$ and $b_{c}^{2}$ is a constant.
As in Section 3, now one can solve for $\omega$ and
$\rho_{\Lambda}$. From (\ref{i1}) and (\ref{i2}) it is easy to
find that
\begin{equation}
\omega=-b_{c}^{2}r-(2+\frac{1}{r_{f}})b_{c}^{2}. \label{r7}
\end{equation}
Then from (\ref{r1}), one finds that the only solution to
$r^{\prime}=0$ is $r=r_{f}$. Note that this result does not
dependent on the choice of $k$, as what can be obtained
from (\ref{r6}) directly. The DE density is found to be%
\begin{equation}
\rho_{\Lambda}=\rho_{\Lambda}^{0}a^{-3(1-b_{c}^{2}-b_{c}^{2}/r_{f})}.
\label{r8}%
\end{equation}
Then from (\ref{f4a}) one finds that%
\[
H^{2}=H_{0}^{2}a^{-3(1-b_{c}^{2}-b_{c}^{2}/r_{f})}\frac{1+r}{1+r_{0}}.
\]

From (\ref{h2}) one finds that the condition $\ddot{a}>0$
requires that%
\[
b_{c}^{2}>\frac{1+r_{0}}{3\left(  r_{0}+2+1/r_{f}\right)  }.
\]
As an example, $r_{f}=0.2$ leads to $b_{c}^{2}>0.06$. To compare
the predictions of this model with low redshift observations, the
distance moduli vs. redshift are plotted in FIG.\ref{fig1}.

As shown in Section 3, given $\tilde{Q}$ and $r$, the characteristic size $L$
and the dynamics of $\omega$ can be determined uniquely. In fact, from
(\ref{r8}) one finds immediately that in HDE models
\begin{equation}
L \propto a^{{3\over 2}(1-b_{c}^{2}-b_{c}^{2}/r_{f})}. \label{r9}
\end{equation}
Moreover, from another point of view, (\ref{r6}) and (\ref{r7})
can also be considered as the consequences of the characteristic
size given by (\ref{r9}).

\subsection{Model 3: Given the characteristic scale of holographic bound $L$}
Similar to the previous subsection, here we consider the situation
where the late time evolution of $r$ to $r_{f}$ can be
approximated by an exponential function of $a$, i.e.
$r=r_{f}(1+\gamma e^{-\lambda a})$ where both $\lambda$ and
$\gamma$ are constants. For simplicity, in the following discussion
we set $\gamma=1$ and thus we have%
\begin{equation}
r=r_{f}(1+e^{-\lambda a}),
\end{equation}
which leads to $r_{f}=r_{0}/(1+e^{-\lambda})$. Nevertheless,
rather than fixing the interaction term $\tilde{Q}$, we consider
in this subsection the HDE model with the characteristic size $L$
being the future event horizon. As discussed in Section 3, from
(\ref{i30}) and
(\ref{h1}), the interaction $\tilde{Q}$ and $\omega$ are found to be%
\begin{align}
\tilde{Q} &  =r-\lambda a(r-r_{f})+\frac{2r}{c\sqrt{1+r}},\nonumber\\
\omega &  =-\frac{1}{3}\left(  1+r+\frac{2\sqrt{1+r}}{c}-\frac{\lambda
ae^{-\lambda a}}{1+e^{-\lambda}}r_{0}\right)  .\label{r10}%
\end{align}

\begin{figure}
[ptb]
\begin{center}
\includegraphics[
height=3.4022in,
width=5.5045in
]
{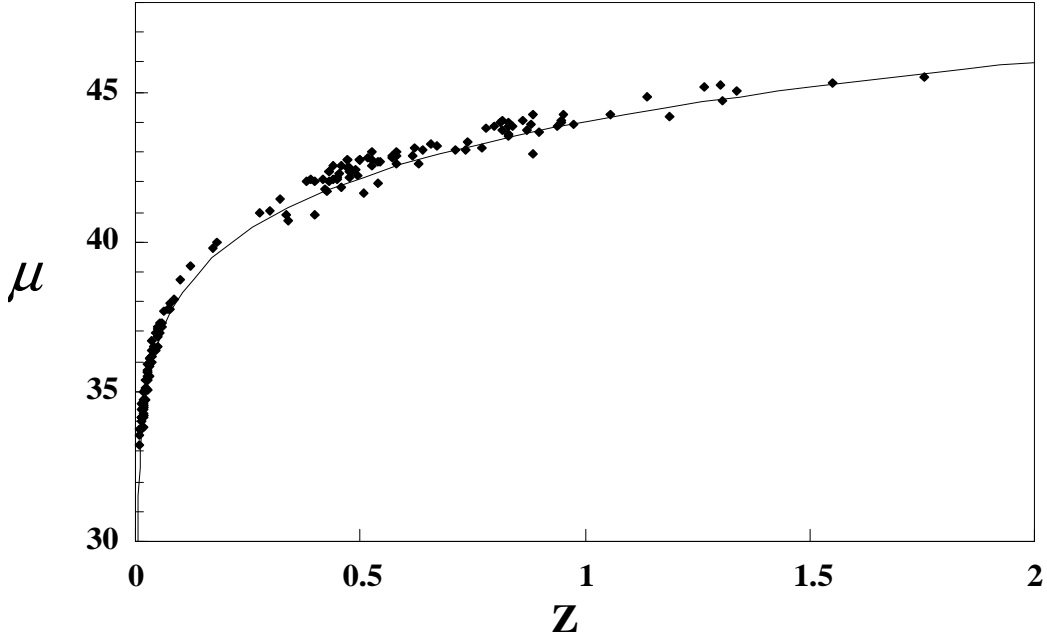}
\caption{Distance moduli vs. redshift plot for model 3 ($c=3$,
$\lambda=0.5$). The data points are from the gold sample of type
Ia supernovae of
\cite{Riess98cb}.}
\label{fig2}
\end{center}
\end{figure}
From (\ref{h2}) and
(\ref{h5}) one finds that $\ddot{a}>0$ and $r^{\prime}<0$ require that
\[
-\frac{2\sqrt{1+r}}{c}<1+r+3\omega<0.
\]
Then from (\ref{r10}) one has
\[
-\frac{2\sqrt{1+r}}{c}<-\frac{2\sqrt{1+r}}{c}+\frac{\lambda ae^{-\lambda a}
}{1+e^{-\lambda}}r_{0}<0.
\]
Since $\lambda ae^{-\lambda a}\leq e^{-1}$,
\[
\frac{\lambda ae^{-\lambda a}}{1+e^{-\lambda}}<\frac{e^{-1}}{1+e^{-\lambda}
}<e^{-1}.
\]
It is easy to check that the above requirement can be satisfied
for any $\lambda$ if $c<2e/r_{0}\simeq15$. Again we can compare
the predictions of this model with low redshift observations, as
shown in FIG.\ref{fig2}.

\begin{figure}
[ptb]
\begin{center}
\includegraphics[
height=3.5319in,
width=5.5391in
]
{
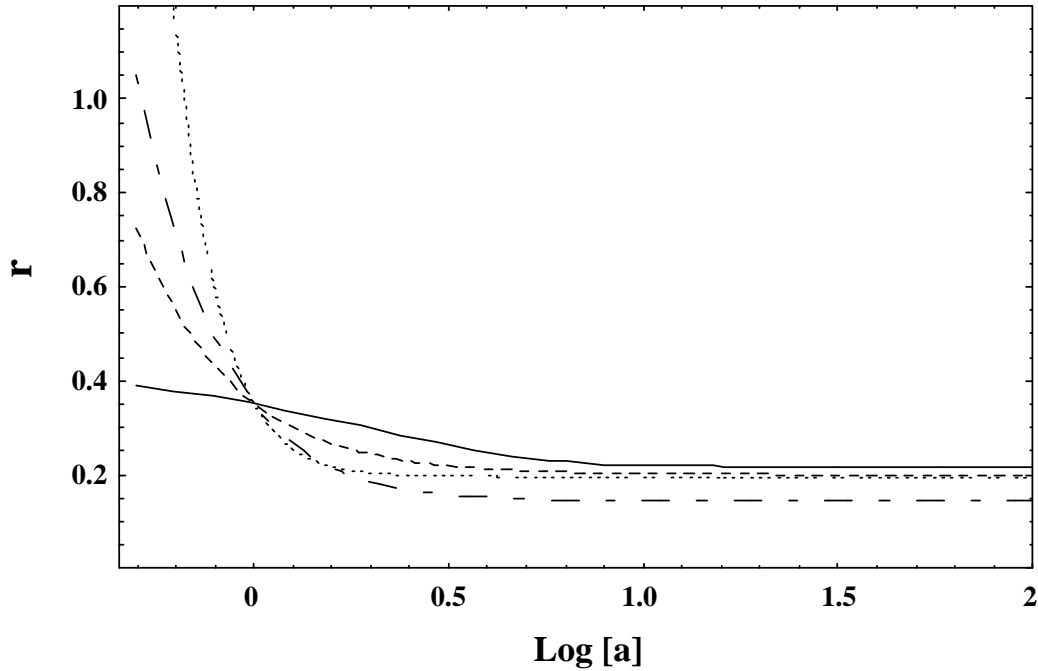}
\caption{$r$ vs. $\log[a]$ (see text for explanations and parameters used for
different curves)}
\label{fig3}
\end{center}
\end{figure}

Moreover, to compare the above three models and the interacting
HDE model discussed at the beginning of this section (see
(\ref{r1a})), the dynamics of $r$ in these models are plotted in
FIG.\ref{fig3} in which the dotted curve corresponds to model 1,
the dashed curve to model 2, the solid curve to model 3 and the
dot-dashed curve to (\ref{r1a}). The parameters used for model 2
and 3 are the same as those for FIG.\ref{fig2} and FIG.\ref{fig3}.
For model 1, the parameters are given in the sentence following
(\ref{r3}). For the curve corresponding to (\ref{r1a}), $c$ and
$b^{2}$ in (\ref{r1a}) are taken to be $1$ and $0.12$,
respectively.

\section{Discussion}

We have presented a 
general formalism of interacting and holographic dark
energy in this paper. Let us summarize the main results as follows. First we
pointed out that in this general formalism both the characteristic size of
holographic bound $L$ and the coupling term of interaction $Q$ for dark energy
are not necessarily fixed as in previous references where these two sorts of
models are separately investigated. 
Given the conditions that our universe is currently accelerating
and the ratio of dark matter to dark energy decreases, we derived the general
relations among the quantities of $L$,$Q$,$\omega$ and $r$ as well as the
constraints on the possible range of these quantities. In particular, the
dynamics of parameters $\omega$ and $r$ are determined by the choice of $L$
and $Q$, thus providing a mechanism to change the evolution of $r$ from that
in standard cosmology such that the coincidence problem may be solved. This is
the main feature of our formalism. Then we proposed three kinds of strategies
to show how the coincidence problem can be alleviated in this context. One
possibility is to have a constant ratio throughout the evolution of the
universe. The second is to have two constant solutions to the ratio $r$ such
that it will run from the maximum constant to the minimum stable one, while
the third and perhaps the most practical one is to have stable constant
solution at the late time but this value is not quite far from our current
observation. Focusing on the third strategy we constructed some models
explicitly and show how this can be implemented by appropriately choosing the
quantities of $L$ and $Q$. In particular our results show that at least in the
low red-shift region these models are well fitted with the current observation.

In all this paper we assume our universe is spatially flat but it is
completely possible to show that the parallel analysis could be 
extended to the spatially closed and hyperbolic universe. We also
expect that the further investigation will provide us a more exact
picture of the dark matter and dark energy by strictly fitting the
observations in high red-shift region.

\section*{Acknowledgement}

This work is partly supported by NSFC (No. 10505011, 10405027, 10205002) and
SRF for ROCS, SEM. B.Hu would like to thank the Institute of Theoretical
Physics of CAS for hospitality during his recent visit.


\begin{thebibliography}{99}                                                     

\bibitem {Riess98cb}A.~G.~Riess \textit{et al.} [Supernova Search Team
Collaboration],
Astron.\ J.\ \textbf{116}, 1009 (1998) [arXiv:astro-ph/9805201].

\bibitem {Perlmutter98np}S.~Perlmutter \textit{et al.} [Supernova Cosmology
Project Collaboration],
Astrophys.\ J.\ \textbf{517}, 565 (1999) [arXiv:astro-ph/9812133].

\bibitem {Bennett03bz}C.~L.~Bennett \textit{et al.},
Astrophys.\ J.\ Suppl.\ \textbf{148}, 1 (2003) [arXiv:astro-ph/0302207].




\bibitem {Spergel03cb}D.~N.~Spergel \textit{et al.} [WMAP Collaboration],
Astrophys.\ J.\ Suppl.\ \textbf{148}, 175 (2003) [arXiv:astro-ph/0302209].




\bibitem {Tegmark03ud}M.~Tegmark \textit{et al.} [SDSS Collaboration],
Phys.\ Rev.\ D \textbf{69}, 103501 (2004) [arXiv:astro-ph/0310723].




\bibitem {Tegmark03uf}M.~Tegmark \textit{et al.} [SDSS Collaboration],
Astrophys.\ J.\ \textbf{606},
702 (2004) [arXiv:astro-ph/0310725].




\bibitem {Weinberg88cp}S.~Weinberg,
Rev.\ Mod.\ Phys.\ \textbf{61}, 1 (1989).




\bibitem {Carroll00fy}S.~M.~Carroll,
Living Rev.\ Rel.\ \textbf{4}, 1 (2001) [arXiv:astro-ph/0004075].




\bibitem {Peebles02gy}P.~J.~E.~Peebles and B.~Ratra,
Rev.\ Mod.\ Phys.\ \textbf{75}, 559 (2003)
[arXiv:astro-ph/0207347].




\bibitem {coincidence}P.J. Steinhardt, in \emph{Critical Problems in Physics},
edited by V.L. Fitch and and D.R. Marlow (Princeton University Press,
Princeton, NJ, 1997);\newline L.P. Chimento, S.A. Jakubi, and D. Pav\'{o}n,
Phys. Rev. D \textbf{62}, 063508 (2000);\newline\textit{ibid.} \textbf{67},
087302 (2003).

\bibitem {Kolb05me}E.~W.~Kolb, S.~Matarrese, A.~Notari and A.~Riotto,
arXiv:hep-th/0503117.




\bibitem {Kolb05da}E.~W.~Kolb, S.~Matarrese and A.~Riotto,
arXiv:astro-ph/0506534.




\bibitem {Ishibashi05sj}A.~Ishibashi and R.~M.~Wald,
Class.\ Quant.\ Grav.\ \textbf{23}, 235 (2006), arXiv:gr-qc/0509108.




\bibitem {Kolb05ze} E.~W.~Kolb, S.~Matarrese and A.~Riotto,
arXiv:astro-ph/0511073.




\bibitem {Cohen98zx}A.~G.~Cohen, D.~B.~Kaplan and A.~E.~Nelson,
Phys.\ Rev.\ Lett.\ \textbf{82}, 4971 (1999) [arXiv:hep-th/9803132].




\bibitem {Hsu04ri}S.~D.~H.~Hsu,
Phys.\ Lett.\ B \textbf{594}, 13 (2004) [arXiv:hep-th/0403052].




\bibitem {Li04rb}M.~Li,
Phys.\ Lett.\ B \textbf{603}, 1 (2004) [arXiv:hep-th/0403127].




\bibitem {Amendola99er} L.~Amendola,  ``Coupled quintessence,''
Phys.\ Rev.\ D \textbf{62}, 043511 (2000)  [arXiv:astro-ph/9908023].




\bibitem {Chimento03ie} L.~P.~Chimento, A.~S.~Jakubi, D.~Pavon and
W.~Zimdahl,
Phys.\ Rev.\ D \textbf{67}, 083513 (2003)  [arXiv:astro-ph/0303145].
G.~Olivares, F.~Atrio-Barandela and D.~Pavon,
Phys.\ Rev.\ D \textbf{71}, 063523 (2005) [arXiv:astro-ph/0503242].




\bibitem {Mangano02gg} G.~Mangano, G.~Miele and V.~Pettorino,
Mod.\ Phys.\ Lett.\ A \textbf{18}, 831 (2003)  [arXiv:astro-ph/0212518].




\bibitem {Farrar03uw} G.~R.~Farrar and P.~J.~E.~Peebles,
Astrophys.\ J.\ \textbf{604}, 1 (2004)  [arXiv:astro-ph/0307316].


\bibitem {Pavon05yx}D.~Pavon and W.~Zimdahl,
arXiv:gr-qc/0505020.


\bibitem {holoref}Q.~G.~Huang and M.~Li,
JCAP \textbf{0408}, 013 (2004) [arXiv:astro-ph/0404229]. Q.~G.~Huang and
Y.~G.~Gong,
JCAP \textbf{0408}, 006 (2004) [arXiv:astro-ph/0403590]. Y.~G.~Gong, B.~Wang
and Y.~Z.~Zhang,
Phys.\ Rev.\ D \textbf{72}, 043510 (2005) [arXiv:hep-th/0412218]. X.~Zhang,
Int.\ J.\ Mod.\ Phys.\ D \textbf{14}, 1597 (2005) [arXiv:astro-ph/0504586].
Z.~Y.~Huang, B.~Wang, E.~Abdalla and R.~K.~Su,
arXiv:hep-th/0501059. Y.~S.~Myung,
Mod.\ Phys.\ Lett.\ A \textbf{20}, 2035 (2005) [arXiv:hep-th/0501023].
Y.~Gong,
Phys.\ Rev.\ D \textbf{70}, 064029 (2004) [arXiv:hep-th/0404030].
Y.~S.~Myung,
Phys.\ Lett.\ B \textbf{610}, 18 (2005) [arXiv:hep-th/0412224]. H.~Kim,
H.~W.~Lee and Y.~S.~Myung,
arXiv:hep-th/0501118. Y.~S.~Myung,
arXiv:hep-th/0502128. Y.~G.~Gong and
Y.~Z.~Zhang,
arXiv:hep-th/0505175.




\bibitem {interef}L.~Amendola and D.~Tocchini-Valentini,
Phys.\ Rev.\ D \textbf{64}, 043509 (2001) [arXiv:astro-ph/0011243]. W.~Zimdahl
and D.~Pavon,
Phys.\ Lett.\ B \textbf{521}, 133
(2001) [arXiv:astro-ph/0105479]. R.~G.~Cai and A.~Wang,
JCAP \textbf{0503}, 002 (2005) [arXiv:hep-th/0411025]. W.~Zimdahl,
arXiv:gr-qc/0505056. L.~Amendola,
Phys.\ Rev.\ D \textbf{62}, 043511 (2000) [arXiv:astro-ph/9908023].
J.~P.~Uzan,
Phys.\ Rev.\ D \textbf{59}, 123510 (1999) [arXiv:gr-qc/9903004]. F.~Perrotta,
C.~Baccigalupi and S.~Matarrese,
Phys.\ Rev.\ D \textbf{61}, 023507 (2000) [arXiv:astro-ph/9906066].
A.~P.~Billyard and A.~A.~Coley,
Phys.\ Rev.\ D \textbf{61}, 083503 (2000) [arXiv:astro-ph/9908224].
V.~Faraoni,
Phys.\ Rev.\ D \textbf{62}, 023504 (2000) [arXiv:gr-qc/0002091].
L.~P.~Chimento, A.~S.~Jakubi and D.~Pavon,
Phys.\ Rev.\ D \textbf{62}, 063508 (2000) [arXiv:astro-ph/0005070]. T.~Chiba,
Phys.\ Rev.\ D \textbf{64}, 103503 (2001) [arXiv:astro-ph/0106550]. A.~Gromov,
Y.~Baryshev and P.~Teerikorpi,
Astron.\ Astrophys.\ \textbf{415}, 813 (2004) [arXiv:astro-ph/0209458].
M.~B.~Hoffman,
arXiv:astro-ph/0307350. M.~Axenides and K.~Dimopoulos,
JCAP \textbf{0407}, 010 (2004) [arXiv:hep-ph/0401238]. W.~Zimdahl and
D.~Pavon,
Gen.\ Rel.\ Grav.\ \textbf{35}, 413 (2003) [arXiv:astro-ph/0210484].
W.~Zimdahl and D.~Pavon,
Gen.\ Rel.\ Grav.\ \textbf{36}, 1483 (2004) [arXiv:gr-qc/0311067]. D.~Pavon,
S.~Sen and W.~Zimdahl,
JCAP \textbf{0405}, 009 (2004) [arXiv:astro-ph/0402067]. N.~Dalal,
K.~Abazajian, E.~Jenkins and A.~V.~Manohar,
Phys.\ Rev.\ Lett.\ \textbf{87}, 141302 (2001) [arXiv:astro-ph/0105317].
L.~Amendola, M.~Gasperini and F.~Piazza,
arXiv:astro-ph/0407573. E.~Majerotto, D.~Sapone and
L.~Amendola,
arXiv:astro-ph/0410543. R.~J.~Scherrer,
Phys.\ Rev.\ D \textbf{71}, 063519 (2005) [arXiv:astro-ph/0410508].
P.~P.~Avelino,
Phys.\ Lett.\ B \textbf{611}, 15 (2005) [arXiv:astro-ph/0411033].




\bibitem {Szydlowski05ph} M.~Szydlowski,
Phys.\ Lett.\ B \textbf{632}, 1 (2006)  [arXiv:astro-ph/0502034].




\bibitem {Wang05jx}B.~Wang, Y.~g.~Gong and E.~Abdalla,
Phys.\ Lett.\ B \textbf{624}, 141 (2005) [arXiv:hep-th/0506069].


W.~Zimdahl and D.~Pavon,
Phys.\ Lett.\ B \textbf{521}, 133 (2001) [arXiv:astro-ph/0105479].


\end{thebibliography}
\end{document}